\documentclass[english,groupedaddress, superscriptaddress,aps,pre,10pt, nofootinbib, notitlepage]{revtex4-1}
\usepackage{hyperref}
\usepackage{geometry}
\usepackage{bigints}
\usepackage{graphicx}
\usepackage{mathtools}
\usepackage{amsmath}
\usepackage{amssymb}

\DeclarePairedDelimiter\abs{\lvert}{\rvert}%
\DeclarePairedDelimiter\norm{\lVert}{\rVert}%

\usepackage{color}
\definecolor{blue}{rgb}{0,0,1}
\definecolor{black}{rgb}{0,0,0}
\newcommand{\revise}[1]{\textcolor{black}{#1}}
\definecolor{dgreen}{rgb}{0,0.5,0}

\definecolor{dred}{rgb}{0.5,0,0}
\definecolor{dyellow}{rgb}{0.75,0.75,0}

\makeatletter
\let\oldabs\abs
\def\abs{\@ifstar{\oldabs}{\oldabs*}}
\let\oldnorm\norm
\def\norm{\@ifstar{\oldnorm}{\oldnorm*}}
\makeatother

\makeatletter
\usepackage{enumerate}

\usepackage[caption=false]{subfig}

\begin{document}
\title{Superstatistical analysis of water time series}
\title{Fluctuations in water time series follow superstatistics}
\title{Fluctuations in  and conductivity in rivers follow superstatistics}
\title{Fluctuations of water quality time series in rivers follow superstatistics}

\author{Benjamin Schäfer}
\thanks{Correspondence: benjamin.schaefer@nmbu.no}
\affiliation{School of Mathematical Sciences, Queen Mary University of London, London E1 4NS, United Kingdom}
\affiliation{Faculty of Science and Technology, Norwegian University of Life Sciences, 1432 Ås, Norway}

\author{Catherine M. Heppell}
\affiliation{Queen Mary University of London, School of Geography, Mile End Road, London E1 4NS, UK}

\author{Hefin Rhys}
\affiliation{Flow Cytometry Science Technology Platform, The Francis Crick Institute, London, UK}

\author{Christian Beck}
\affiliation{School of Mathematical Sciences, Queen Mary University of London, London E1 4NS, United Kingdom}

\begin{abstract}
Superstatistics is a general method from nonequilibrium statistical physics which has been applied to a variety of complex systems, ranging from hydrodynamic turbulence to traffic delays and air pollution dynamics. Here, we investigate water quality time series (such as dissolved oxygen concentrations and electrical conductivity) as measured in rivers, and provide evidence that they exhibit superstatistical behaviour.
Our main example are time series as recorded in the river Chess in South East England.  Specifically, we use seasonal detrending and empirical mode decomposition (EMD) to separate trends from fluctuations for the measured data. With either detrending method, we observe heavy-tailed fluctuation distributions, which are well described by a log-normal superstatistics for dissolved oxygen. Contrarily, we find a double peaked non-standard superstatistics for the electrical conductivity data, which we model using two combined $\chi^2$-distributions.
\end{abstract}
\maketitle
\makeatother

\section*{Introduction}

Superstatistical methods, as introduced in \cite{beck-cohen2003,BCS}, provide
a general approach to describe the dynamics of complex nonequilibrium systems with well-separated time scales. These models generate heavy-tailed non-Gaussian distributions by a simple mechanism, namely the
superposition of simpler distributions whose relevant parameters are random variables, fluctuating on a much larger time scale. Originating in turbulence modelling \cite{beck2007}, superstatistics has been applied to many  physical systems, such as plasma physics \cite{livadiotis2017,davis2019}, Ising systems \cite{cheraghalizadeh2021superstatistical}, cosmic ray physics \cite{yalcin2018generalized,smolla2020universal}, self-gravitating systems \cite{ourabah2020}, solar wind
\cite{livadiotis2018generation}, high energy scattering processes \cite{beck2009HighEnergy, sevilla2019stationary,ayala2020},
ultracold gases \cite{rouse2017} and non-Gaussian diffusion processes in small complex systems \cite{chechkin2017brownian, itto2021}. 
Furthermore, the framework has successfully been applied to completely different areas, such as modelling the power-grid frequency \cite{schafer2018non}, wind statistics \cite{weber2019wind}, air pollution \cite{williams2020superstatistical}, bacterial DNA \cite{bogachev2017superstatistical},
financial time series \cite{gidea2018topological,uchiyama2019superstatistics}, rain fall statistics \cite{de2018superstatistical} or train delays \cite{briggs2007modelling}. The overview article \cite{metzler2020superstatistics} provides \revise{a recent introduction to superstatistics and non-Gaussian diffusion}.
In all these cases, an underlying simple distribution, typically Gaussian or exponential, is identified to explain the observed heavy tails of the marginal distributions 
when aggregated with the fluctuating parameter. These tails
often decay with a power law. Note that in astrophysical plasmas the so-called $\kappa$-distributions \cite{livadiotis2017} are a typical example of marginal distributions arising in this context, whereas in statistical physics one uses the so-called $q$-Gaussians \cite{tsallis2009}, with $q$ related to $\kappa$ by $\kappa =1/(q-1)$. Both approaches are equivalent, and form standard examples of distributions generated by the (more general) superstatistical approach.

A common feature of real-world time series is that they consist of some long-term trend or oscillation combined with short-term fluctuations. Consider a time series connected to the environment, such as ambient temperature: This will typically display strong seasonal cycles \cite{kumar2009assessment}. Day-Night cycles add another oscillation, while global warming or other long-term influences, such as deforestation might induce a drift towards higher values. We can decompose the full time series in slower seasonal and drift (trend) terms as well as the short-term fluctuations, using detrending methods. In particular, we consider seasonal detrending, i.e. moving averages, and decomposition via empirical mode decomposition (EMD) \cite{EEMD}, which has recently been shown to disentangle short-term fluctuations from longer-term signals \cite{kampers2020disentangling}. Naively, one would expect the so-extracted short-term fluctuations to follow Gaussian distributions.

In this paper, we analyse environmental time series for the river Chess, which is a river located in South East England and is being actively monitored by a citizen science project \cite{RiverChess_CitizenScience}. Key questions include how urban areas and a local sewage treatment works affect the water quality. 
Many different quantities determine the water quality of a river. Here, we focus on two particular quantities: Dissolved oxygen concentration and electrical conductivity of the river. 
Dissolved oxygen (or just "oxygen" for large parts of the paper) is highly relevant for aquatic life, such as fish, in rivers. Meanwhile, electrical conductivity (abbreviated as "EC" or "conductivity") measures the total dissolved solutes in the water. Thereby, it also measures the impact of humans, e.g. via treated effluent water that is fed into the river. 
For the current paper, we utilize data available from ChessWatch \cite{ChessWatch}, from the four locations  Blackwell Hall (BH) {[}Red{]}, Little Chess (LC) {[}Blue{]}, Latimer Park (LP) {[}Green{]} and Watercress Beds (WB) {[}Purple{]}. 
About twelve months of data collected within the time span of June 2019 to May 2020 are evaluated here. Note that LC and BH are upstream of a sewage treatment works, while LP and WB both are downstream of the sewage treatment works. A detailed discussion on how daily cycles influence EC and how machine learning can be used to predict and understand EC trajectories can be found in a future paper \cite{hess2020}. Our main result of the current paper is that the detrended time series behave in a superstatistical way.

This paper is structured as follows. First, we introduce the data and
discuss the trajectories and empirical probability density functions (PDF) of oxygen and EC. Next, we discuss how daily and seasonal cycles are subtracted from the data to reveal the fluctuations. We then continue to present a short re-cap of superstatistical theory to analyse distributions as generated by a given time series, specifically adapted to our problem here. Finally, we use superstatistical methods to extract long time scales and microscopic distributions  of the fluctuating superstatistical parameter $\beta$ as a function of the detrending parameters. Overall, we find that oxygen fluctuations follow approximately log-normal superstatistics, while EC fluctuations point to a new form of superstatistics with a double-peaked $\beta$-distribution at the LC site. 

\section*{Results}
\subsection*{Trajectories and Probability distributions\label{sec:Trajectories-and-Probability}}
\begin{figure}[ht!]
\begin{centering}
\includegraphics[width=0.95\columnwidth]{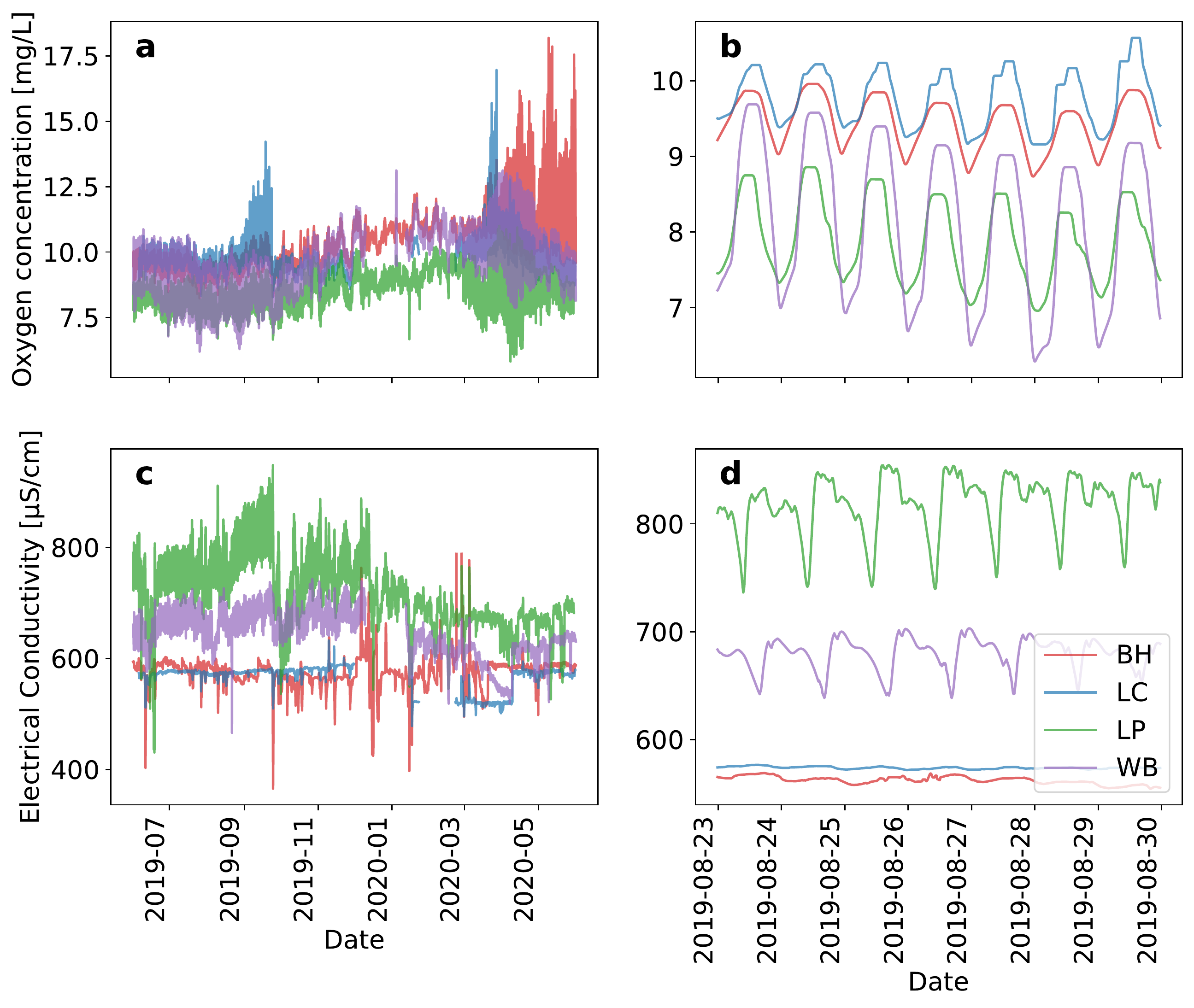}
\par\end{centering}
\caption{Trajectories of the the oxygen concentration (a-b) and the electrical conductivity (c-d). We display both the full time period of available data (a,c) and a one-week extract (b,d), highlighting the daily cycles. \label{fig:Trajectories}}
\end{figure}

To obtain an initial impression of the water quality dynamics, we visualize the trajectories of the oxygen concentration and the electrical conductivity in Fig. \ref{fig:Trajectories}.
Disregarding some large peaks at the BH and LC sites, we observe certain seasonal trends in the oxygen trajectories (Fig. \ref{fig:Trajectories} a), i.e. higher concentrations of oxygen in winter to spring than during the summer. On a shorter time scale, both oxygen and electrical conductivity  show obvious daily cycles at all stations (Fig. \ref{fig:Trajectories} b,d). 
\revise{Electrical conductivity provides a measure of total dissolved solutes in water. Urban streams tend to have higher mean electrical conductivity and major ion concentrations in comparison to their rural counterparts \cite{conway2007impervious,rose2007effects,peters2009effects}, which arises from a combination of point and diffuse pollution sources. Dissolved oxygen content is a critical indicator of river health for biota, and low dissolved oxygen content or strong daily changes in dissolved oxygen will cause harm to many organisms living in chalk streams such as the river Chess \cite{arroita2019twenty,rajwa2015dissolved}.}

\begin{figure}
\begin{centering}
\includegraphics[width=0.95\columnwidth]{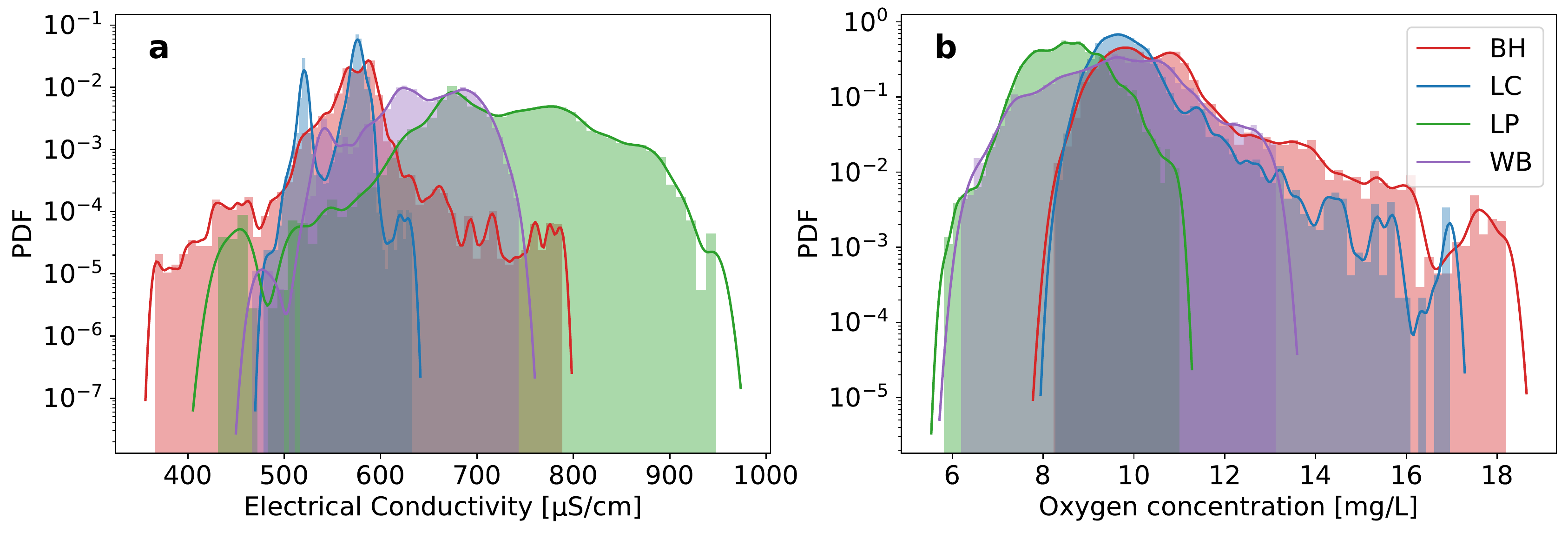}
\par\end{centering}
\caption{Aggregated statistics points to non-Gaussian dynamics. We display the empirical probability density function (PDF) of the electrical conductivity (a) and the oxygen concentration (b). The lines are Gaussian kernel estimates of the empirical PDF. Note the log-scale on the y-axis. \label{fig:Histograms}}
\end{figure}

Intriguingly, the aggregated statistics shows clear deviations from Gaussianity, see  the empirical probability density functions (PDF) of both quantities in Fig. \ref{fig:Histograms}. In particular the sites BH and LC (red and blue) display heavy tails. 
Still, a large portion of the observed variability arises due to daily and seasonal cycles, which we have to subtract from the data before we continue our statistical analysis.

\subsection*{Detrending}
Instead of modelling the full distribution, with its daily and seasonal dynamics, we will describe the fluctuations of the water quality parameters around their respective trend. \revise{Detrending reduces the variability and allows for weak stationarity in time series, thus allowing forecasting with more precision \cite{contreras2020backcasting}.} To carry out the detrending, we first need to separate the full trajectory $F(t)$ into trend and fluctuations (assuming an additive model):

\begin{equation}
    F(t)=\mathrm{Trend}(t)+\mathrm{Fluctuations}(t).
\end{equation} To achieve this separation, we employ two different methods: Seasonal decomposition and EMD. 

Seasonal decomposition applies a moving average to the data with a filtering frequency $f$ to obtain the trend. The deviation between this moving average and the original data is then classified as fluctuations. Technically, we implement it via the python \emph{statsmodels.tsa.seasonal} package \cite{StatsModel} and typically apply $f=6$ hours. 

Alternatively, the empirical mode decomposition (EMD) splits the full trajectory into ordered modes ranging from slowly changing to highly oscillating modes. Similar to a Fourier analysis, summing all modes, it yields the full original data. As has been pointed out recently \cite{kampers2020disentangling}, EMD can be used to disentangle deterministic and stochastic influences. Here, we do the following. All modes $h_i(t)$ summed up form the full dynamics

\begin{equation}
    F(t)=\sum_{i=1}^N h_i(t),
\end{equation}
where $N$ is the total number of modes. Since the lower numbered modes represent the trend, we keep all but the last $m$ modes for the trend and declare the remaining modes as the fluctuations, i.e.
\begin{equation}
    \mathrm{Trend}(t)=\sum_{i=1}^{N-m} h_i(t),
\end{equation}
\begin{equation}
    \mathrm{Fluctuations}(t)=\sum_{i=N-m+1}^{N} h_i(t).
\end{equation}
Technically, we implement the EMD via the PyEMD package \cite{pyemd} and chose $m=2$ for most cases.

\begin{figure}
\begin{centering}
\includegraphics[width=0.95\columnwidth]{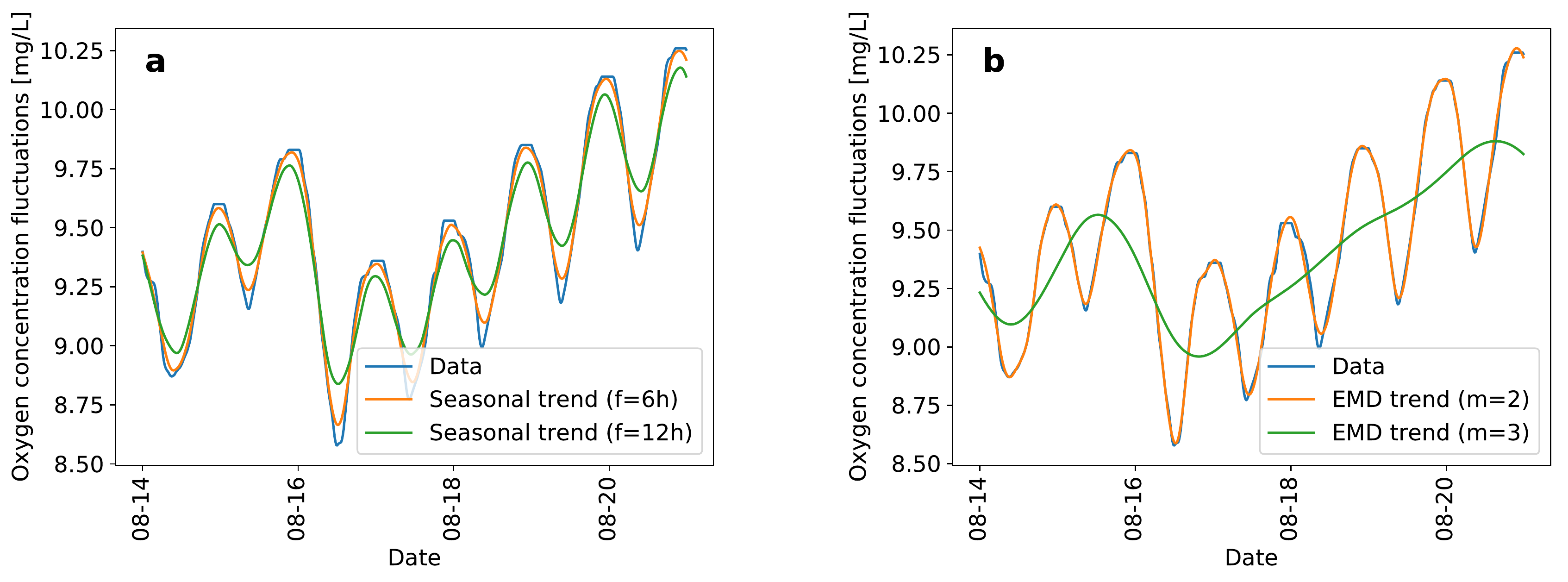}
\par\end{centering}
\caption{Illustration of the data detrending. We apply seasonal detrending (a) or detrending via EMD (b). The data (blue) is best approximated by a filtering frequency of $f=6h$ and dropping $m=2$ modes respectively (orange). Choosing a larger $f$ or $m$ oversimplifies the dynamics (green), while smaller settings would overfit the noise. Here, we plot a one week extract of the oxygen trajectory for the BH measurement site. Note that the EMD is still carried out on the full data set, as the number of modes per individual week would vary otherwise.  \label{fig:Detrending_illustration}}
\end{figure}

Both detrending procedures are demonstrated in Fig. \ref{fig:Detrending_illustration} using oxygen concentrations from the BH measurement site. The orange curves, corresponding to a filtering frequency of $f=6h$ or dropping $m=2$ modes describes the trend of the data well, while preserving  short-time scale fluctuations. These parameter settings are a compromise between barely capturing any trend (green curves) and overfitting (essentially reproducing the blue data). We will later study the effect of the detrending parameters on the superstatistical results systematically. 
\revise{With the data separated into trend and fluctuations, let us now continue to investigate the fluctuation statistics using a superstatistical approach.}

\subsection*{Superstatistical time series analysis}

The basic idea of superstatistics \cite{beck-cohen2003,BCS} is the concept that a longer time series with a complicated and often heavy-tailed probability distribution is indeed an aggregation of many shorter time series, each giving rise to a simple, non-heavy-tailed distribution. Superstatistical methods have been successfully applied to many different types
of complex systems \cite{beck-cohen2003}--\cite{metzler2020superstatistics}.
As a first step of superstatistical time series analysis, we will have to extract a long time scale $T$ on which we locally observe simple distributions. Assume we know that locally,
in shorter time slices, the time series is approximately Gaussian
distributed. In this case, the kurtosis of a local snapshot should be  $\kappa_\text{Gaussian}=3$. In contrast, the fully aggregated time series will display a much higher kurtosis $\kappa$. To determine $T$, we test different time window sizes $\Delta t$ and compute the local average kurtosis \cite{BCS} as 
\begin{equation} \label{Eq4}
\bar{\kappa}\left(\Delta t\right)=\frac{1}{t_{max} - \Delta t}\int_{0}^{t_{max}-\Delta t} dt_{0}\frac{\langle\left(u-\bar{u}\right)^{4}\rangle_{t_{0},\Delta t}}{\langle\left(u-\bar{u}\right)^{2}\rangle_{t_{0},\Delta t}^{2}},
\end{equation}
where  $t_{max}$ is the length of the time series and $\langle\dots\rangle_{t_0, \Delta t}$ is the expectation for the time slice of length $\Delta t$ starting at $t_0$. The long time scale is then \revise{assumed} as $\bar{\kappa}(T)=\kappa_\text{Gaussian}$, i.e. the average kurtosis of windows of length $T$ has a Gaussian kurtosis $\bar{\kappa}(T)=3$.
After determining $T$, we can split the time series in several samples, each of length $T$ and thereby obtain a collection of approximately local Gaussian distributions, each with a different inverse variance parameter $\beta$. If these $\beta$ themselves follow a $\chi^2$-distribution
\begin{equation}
f(\beta)=\frac{1}{\Gamma\left(\frac{n}{2}\right)}\left( \frac{n}{2 \beta} \right)^{\frac{n}{2}} \beta^{\frac{n}{2}-1}e^{-\frac{n\beta}{2\beta_0}}, \label{fbeta}
\end{equation}
with $n$ being the degrees of freedom for the distribution and $\beta_0$ the  mean of $\beta$, we then analytically obtain a $q$-Gaussian for the aggregated statistics \cite{beck2001,BCS}.
Alternatively, the $\beta$-distribution might be well described by 
some other distribution, such as an inverse $\chi^2$ or log-normal distribution. In this case the marginal
distribution obtained by integrating over $\beta$ is different (though often,
in good approximation, well-approximated by a $q$-Gaussian). A given time series is then said to follow a $\chi^2$, an inverse $\chi^2$ or a log-normal superstatistics, depending on what the actual distribution of $\beta$ is. 
\revise{As superstatistics was originally derived for temperature fluctuations, $\beta$ is often interpreted as an inverse temperature \cite{uchiyama2019superstatistics}, related to the local kinetic energy in the system. But in general it is just a fluctuating inverse variance parameter of a given time series.}

For a generic superstatistics, we expect to observe in good approximation $q$-Gaussian probability density functions, which are given as 
\begin{equation}
    p(q,b, \mu)=\frac{\sqrt{b}}{C_q}\left(1+(1-q)(-b (x-\mu)^2) \right)^\frac{1}{1-q},
\end{equation}
where $C_q$ is the normalization constant, $\mu$ is a shift parameter, $q$ is a shape parameter, also  known as the entropic index, and $b$ is a scale parameter proportional to the expectation $\langle \beta \rangle$ as formed with
the distribution given in eq.~(\ref{fbeta}). For $q\to1$, $q$-Gaussians become Gaussian distributions with variance $\frac{1}{2 b}$. For a specialized book on the applications of $q$-statistics in
water engineering, see \cite{singh2016}.
\revise{Note that the superstatistical distributions described here may arise from a Gaussian process if such a process has a time-dependent standard deviation, i.e. displays a superposition of simple Gaussian distributions, in the long term.}

\revise{While the long time scale $T$ describes the time scale on which the underlying stochastic process changes, the short time scale $\tau$ gives the time for the system to relax towards its (local) equilibrium. It is defined by evaluating the decaying autocorrelation of the original time series, approximated by $c \sim \text{e}^{(-t/ \tau )}$. To ensure that the system can always relax to its new equilibrium, we have to assume $\tau\ll T$ for the superstatistical approach to hold. We validate this in the Supplemental Material.}

\subsection*{Superstatistical analysis for the river Chess\label{sec:Superstatistical-analysis}}

\begin{figure}
\begin{centering}
\includegraphics[width=0.95\columnwidth]{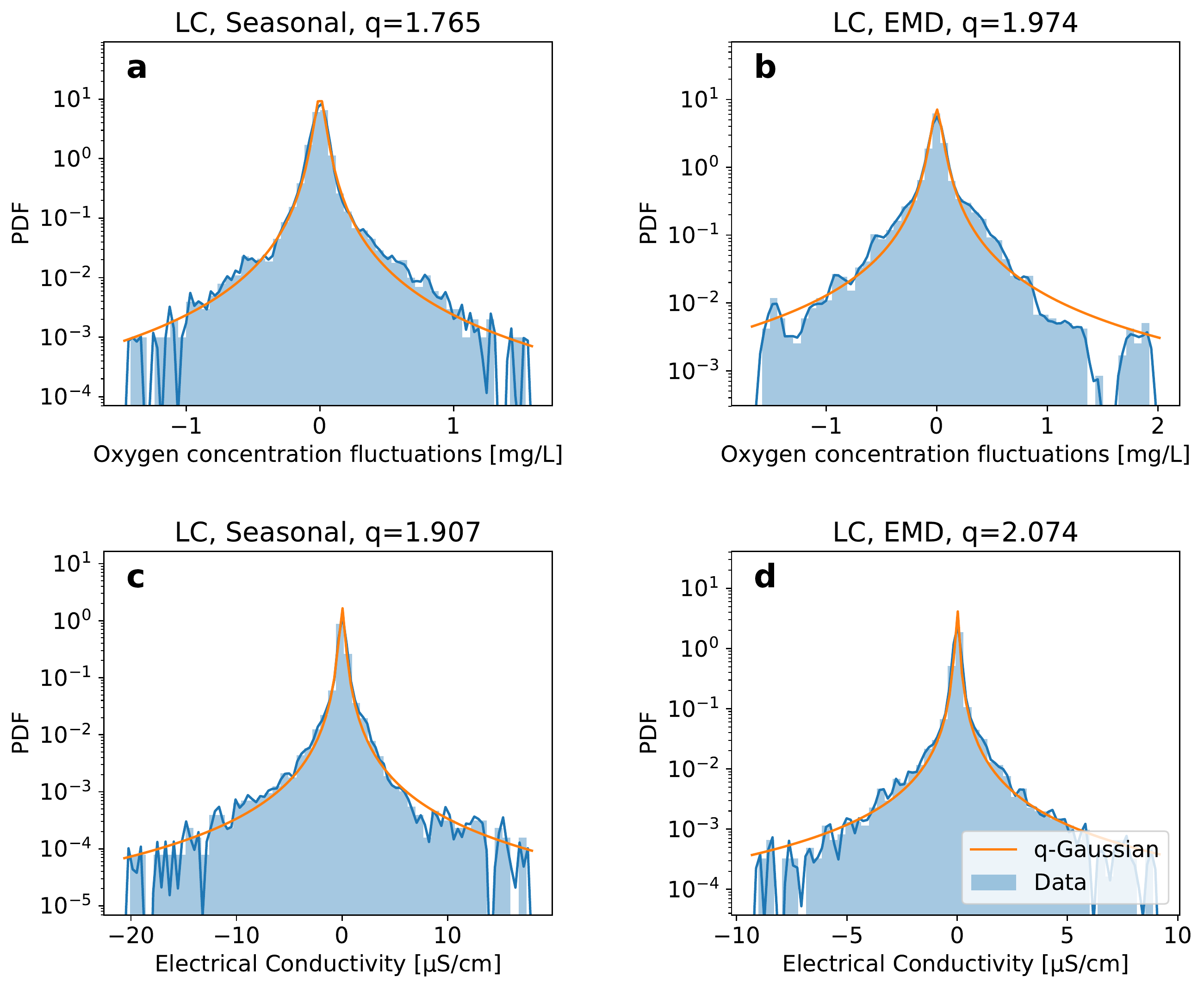}
\par\end{centering}
\caption{Detrending of the data reveals non-Gaussian fluctuations, approximated by $q$-Gaussians in both cases. We plot the empirical probability density functions (PDF) of detrended oxygen concentrations (a-b) and electrical conductivity (c-d). Regardless whether the detrending is carried out via seasonal detrending (a,c) or EMD (b,d) leads to these non-Gaussian distributions, which are well-approximated by $q$-Gaussian distributions. The blue lines are Gaussian kernel estimates of the empirical PDF. \revise{The orange fits of $q$-Gaussians were obtained via maximum likelihood estimation (MLE), see code for details.} \label{fig:Histograms_detrended}}
\end{figure}

With the data detrended and the superstatistical foundations laid out, let us investigate the fluctuations in the two time series for the river Chess data (oxygen and electrical conductivity). First, we note that the detrending of either water quality parameter leaves us with a highly non-Gaussian distribution, which is well-captured by a $q$-Gaussian distribution, see Fig. \ref{fig:Histograms_detrended}. 

\begin{figure}
\begin{centering}
\includegraphics[width=0.5\columnwidth]{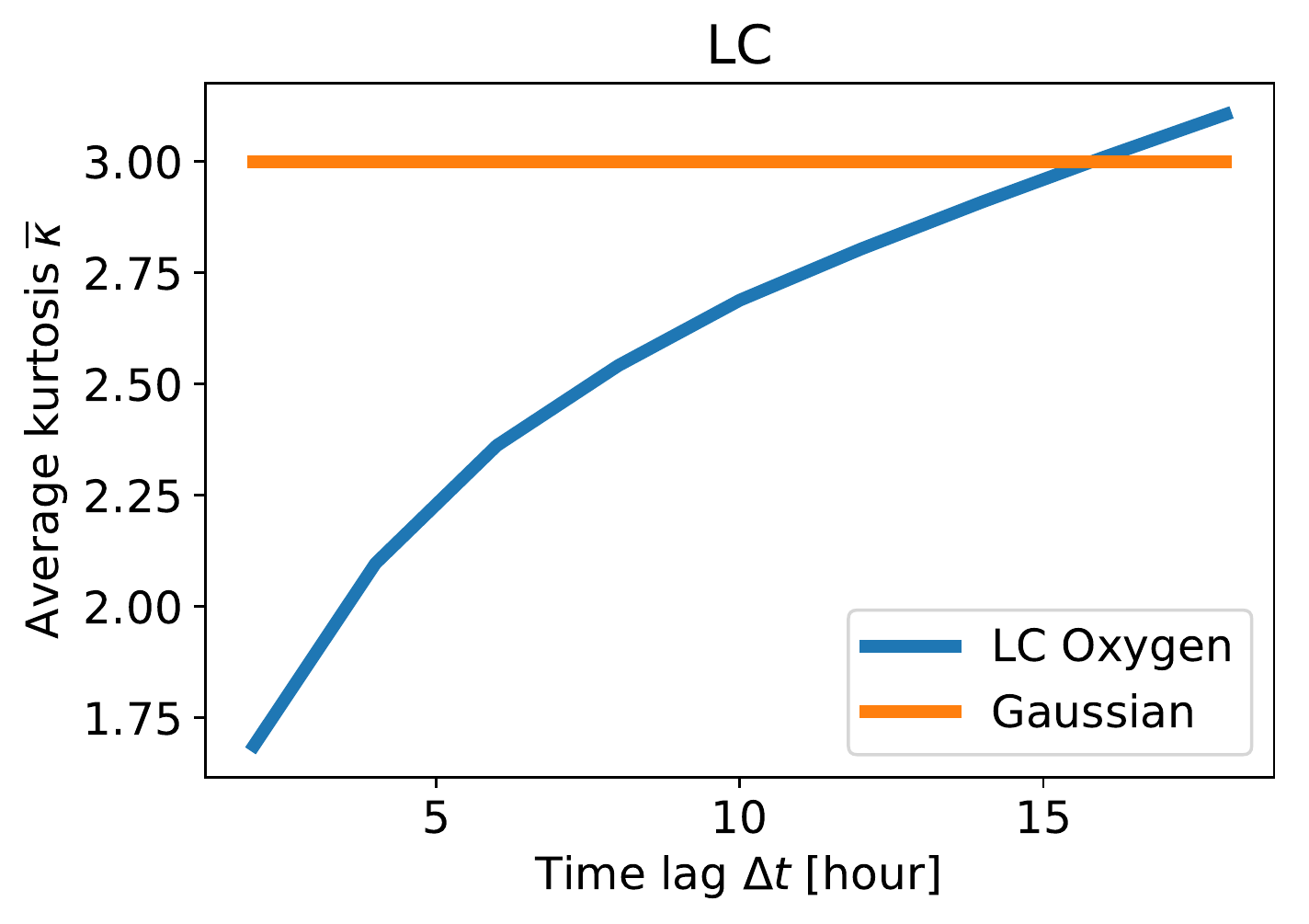}
\par\end{centering}
\caption{The long time scale is determined using the average kurtosis. Specifically, we display the average kurtosis $\bar{\kappa}$ as a function of the time window $\Delta t$ and determine $T$ from the condition $\kappa(T)=3$. Assuming Gaussian distributions locally in a window of length $T$, they have kurtosis 3, whatever their variance. In this way, for the LC site displayed here, we obtain $T\approx16h$. \label{fig:Long_time_scale}}
\end{figure}

To investigate how these non-Gaussian distributions could arise, we continue with the superstatistical ansatz: Let us assume that the non-Gaussian fluctuations arise from local Gaussian distributions. If this was the case, we could extract a long time scale $T$ on which the distribution is locally a Gaussian distribution. We determine this long scale as the time window for which the average kurtosis $\bar{\kappa}$ of the data is $\bar{\kappa}(T)=\kappa_{\text{Gauss}}=3$, see Fig. \ref{fig:Long_time_scale}. For the LC measurement site, using seasonal detrending and investigating  oxygen concentrations, we observe a long time scale of $T_{\text{LC}}\approx16\,h$.

\begin{figure}
\begin{centering}
\includegraphics[width=0.95\columnwidth]{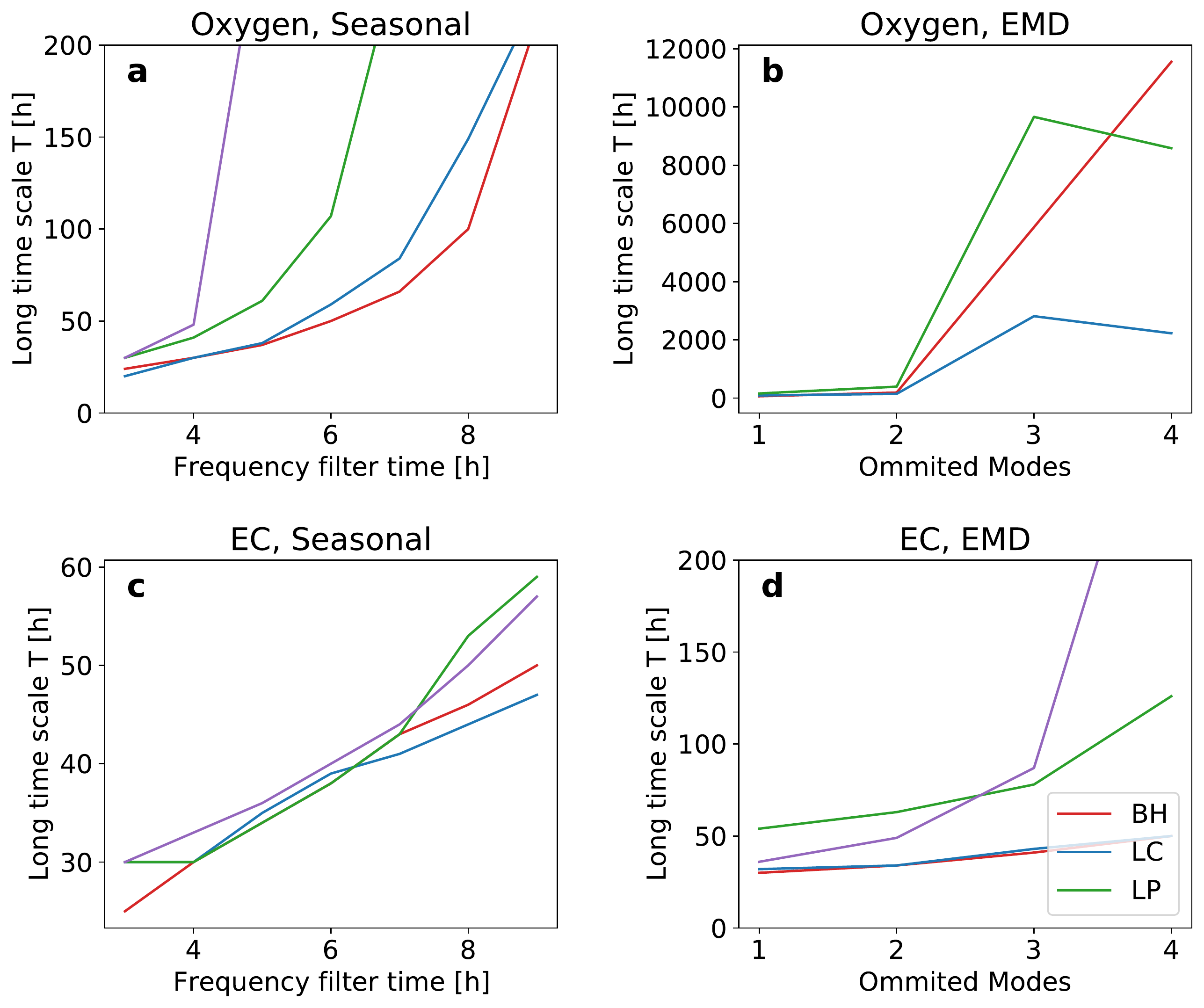}
\par\end{centering}
\caption{Long time scales $T$ scale almost linearly with the detrending parameters before the description breaks down. We plot the obtained long time scale $T$ for the  detrended oxygen concentrations (a-b) and electrical conductivity (c-d), considering both detrending via seasonal detrending (a,c) or EMD (b,d). If the number of omitted modes $m$ or the filtering frequency $f$ are set too high, the average kurtosis always remains below $\kappa^\text{Gaussian}=3$ and hence no time scale $T$ is determined in this case.\label{fig:Long_time_scale_scaling}}
\end{figure}

Let us continue this investigation more systematically. Namely, as pointed out above, the detrending method and detrending  parameter (filtering frequency $f$ and number of omitted modes $m$) will likely influence the superstatistics and thereby the long time scale. Hence, we visualize this dependency for both methods and both quantities in Fig. \ref{fig:Long_time_scale_scaling}.
Apparently, the long time scale scales approximately linearly with the detrending parameter in a certain parameter range. Then, when the detrending parameter is increased too much (e.g. at $f>4h$ for seasonal detrending and oxygen or $m>2$ for EMD and oxygen), the long time scale suddenly increases dramatically. This can be explained as follows: The influence of the specific detrending parameter on the long time scale is moderate as long as the derived fluctuation distribution is heavy-tailed. If too many modes are attributed to the fluctuations (large $m$) or too high frequencies are used (high filter frequency $f$), then the fluctuation distributions might only barely be heavy tailed (high $T$) or display a platykurtic behavior, i.e. a kurtosis $\kappa<3$. Based on the results seen here, we are confident that a filtering frequency of $f=6h$ and attributing $m=2$ modes to the fluctuations yields  solid results for as many cases as possible. The special case of the WB site, which would require $f\leq 4h$ is thereby not included to avoid overfitting at the other sites. With the method established, let us carry out two consistency checks: Snapshots and $\beta$-distribution.

\begin{figure}
\begin{centering}
\includegraphics[width=0.95\columnwidth]{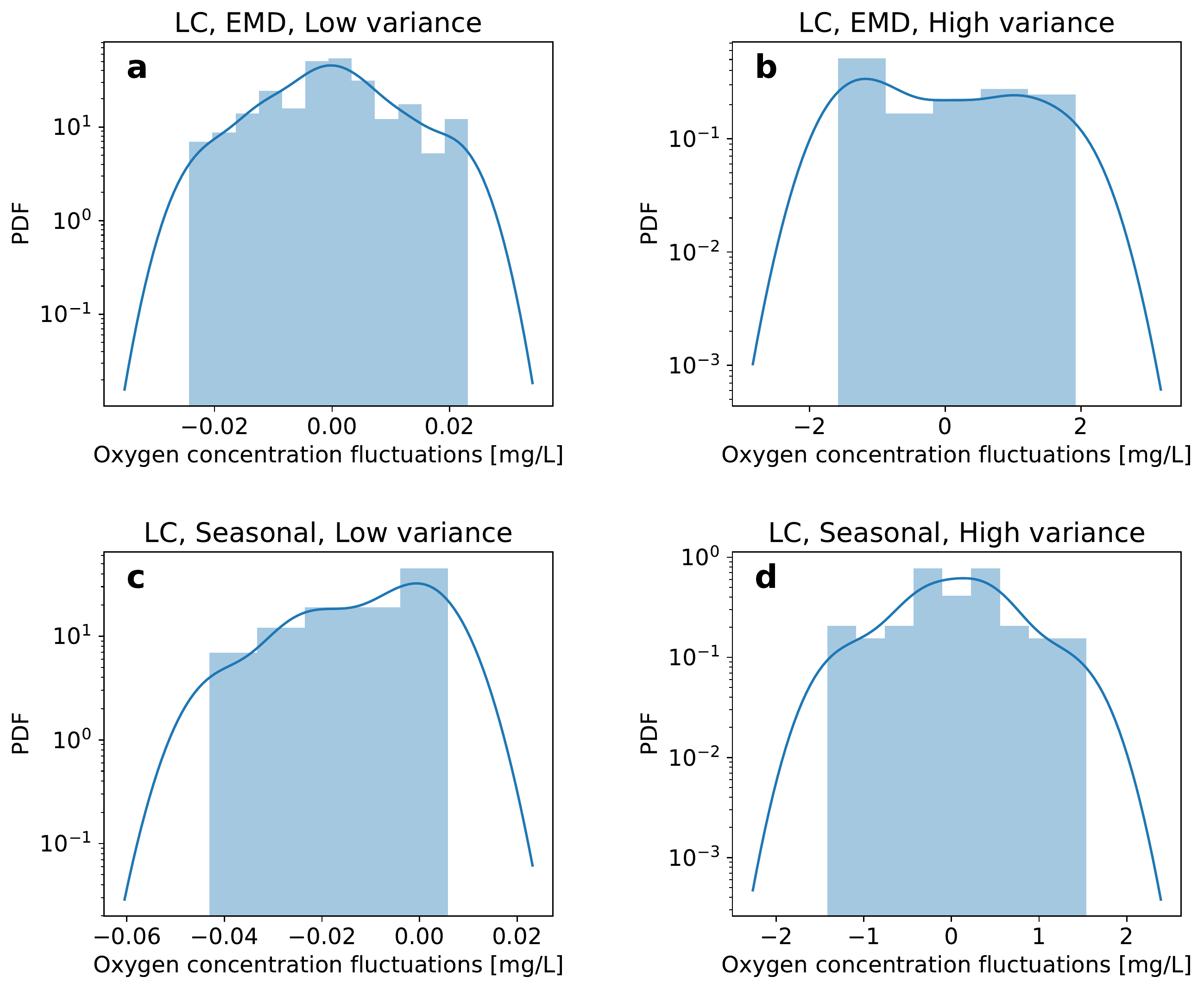}
\par\end{centering}
\caption{Local snapshots of length $T$ are approximated by Gaussian distributions. We consider both EMD detrending (a-b) and Seasonal detrending (c-d) and display for both cases a window of length $T$, selecting cases with lowest variance (a,c) and the highest variance (b,d). All plots are for the LC site data.  The figure illustrates how strongly the local variance fluctuates. The blue lines are Gaussian kernel estimates of the empirical PDF. \label{fig:Snapshots}}
\end{figure}

First, we inspect snapshots of the fluctuation trajectory of length $T$. According to the superstatistical approach, these local snapshots should follow a Gaussian distribution. Indeed, inspecting the plots in Fig. \ref{fig:Snapshots}, we observe approximately Gaussian distributions. Note that the long time scale here is of the order of 10-100 hours and the data has 15 minute resolution, i.e each local snapshot contains $\sim$ 100-1000 measurements.

\begin{figure}
\begin{centering}
\includegraphics[width=0.95\columnwidth]{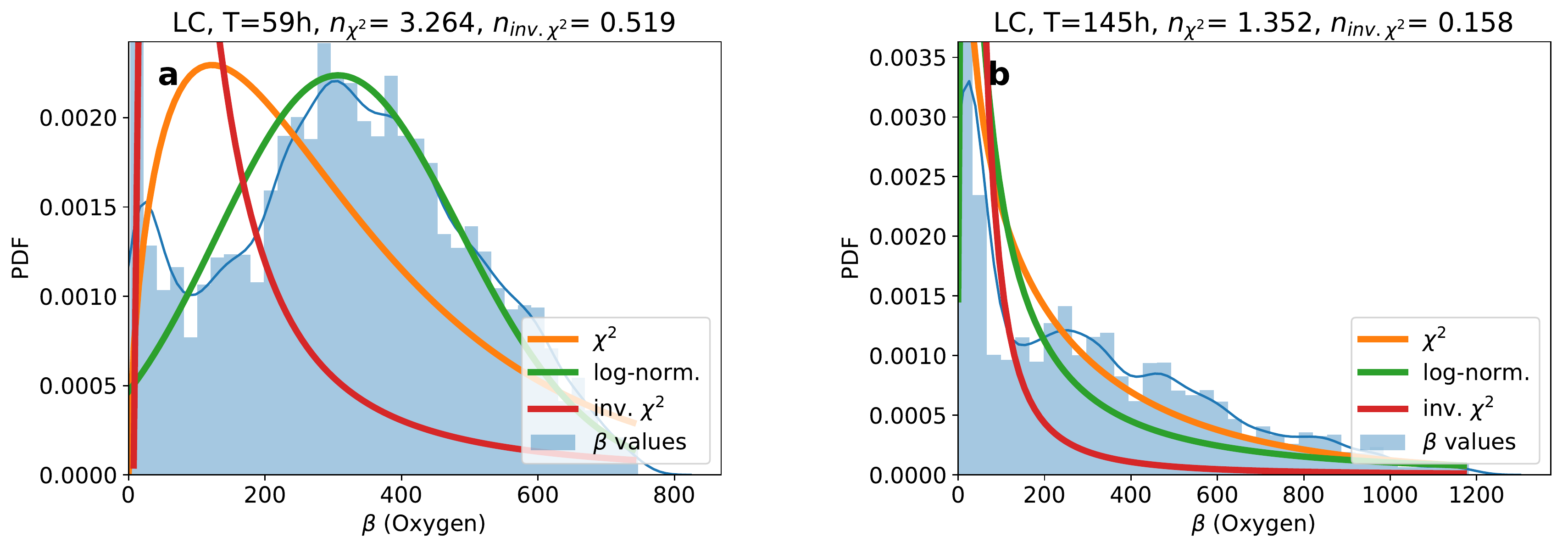}
\par\end{centering}
\caption{The extracted $\beta$-distribution of the oxygen concentration fluctuations is well approximated by a log-normal fit. Here, we assume local Gaussian distributions (with fluctuating variance in each time slice) and investigated the LC measurement site, considering both seasonal detrending (a) and EMD (b). The blue lines are Gaussian kernel estimates of the empirical PDF.\label{fig:Beta_dist_Ox}}
\end{figure}

Finally, we compute the distribution of the effective damping to noise ratio $\beta$. The superstatistical hypothesis implies that the observed heavy tails (fitted $q$-Gaussian-like distributions in Fig.\ref{fig:Histograms_detrended}) arise  either exactly from $\chi^2$,  or approximately from inverse $\chi^2$ or log-normal distributions of $\beta$. Here we observe something very interesting: 
While the $\beta$-distributions of the Oxygen fluctuations are well approximated via log-normal or alternatively $\chi^2$ distributions (Fig. \ref{fig:Beta_dist_Ox}), the $\beta$-distributions of the electrical conductivity fluctuations do follow a very different type of distribution (Fig. \ref{fig:Beta_Mixture}).
While the $\beta$-distribution for oxygen is single-peaked, the one of the electrical conductivity displays two peaks: One close to zero and one at larger values of $\beta$.  These distributions with two peaks are  somewhat unusual distributions, typically not encountered in the standard superstatistics formalism. They provide something new and are specific to the data analysed here. Remember that the electric conductivity is heavily influenced by human influences, such as the outflow of the sewage treatment works, which could be the deeper reason for the observed unusual behaviour: The single-peaked $\beta$-distributions at the LP and BW sites could arise due to human influence, while the double-peaked $\beta$ distributions at the LC site might hint at complex natural processes, e.g. interaction of rainfall events or the flora and fauna with the conductivity fluctuations.



\subsubsection*{Mixture of $\chi^2$-distributions}
Let us search for a suitable description of the double-peaked $\beta$-distribution observed for electrical conductivity. 
As a simple extension of a single $\chi^2$-distribution, we propose to use a mixture distribution of two $\chi^2$-distributions
\begin{equation}
    f(\beta)=W f_{\chi^2}(\beta, n_{\chi_1}, \beta_0) + (1-W) f_{\chi^2}(\beta, n_{\chi_2}, \beta_0),
\end{equation}
i.e. the full $\beta$-distribution is composed as a sum of two $\chi^2$-distributions, sharing a $\beta_0$ parameter (originally the mean $\beta$) and   each having its own degree of freedom $n_{\chi_1}$ and $n_{\chi_2}$. Both distributions are weighted by the weight constant $W$, which ranges from 0 to 1. 

\begin{figure}
\begin{centering}
\includegraphics[width=0.95\columnwidth]{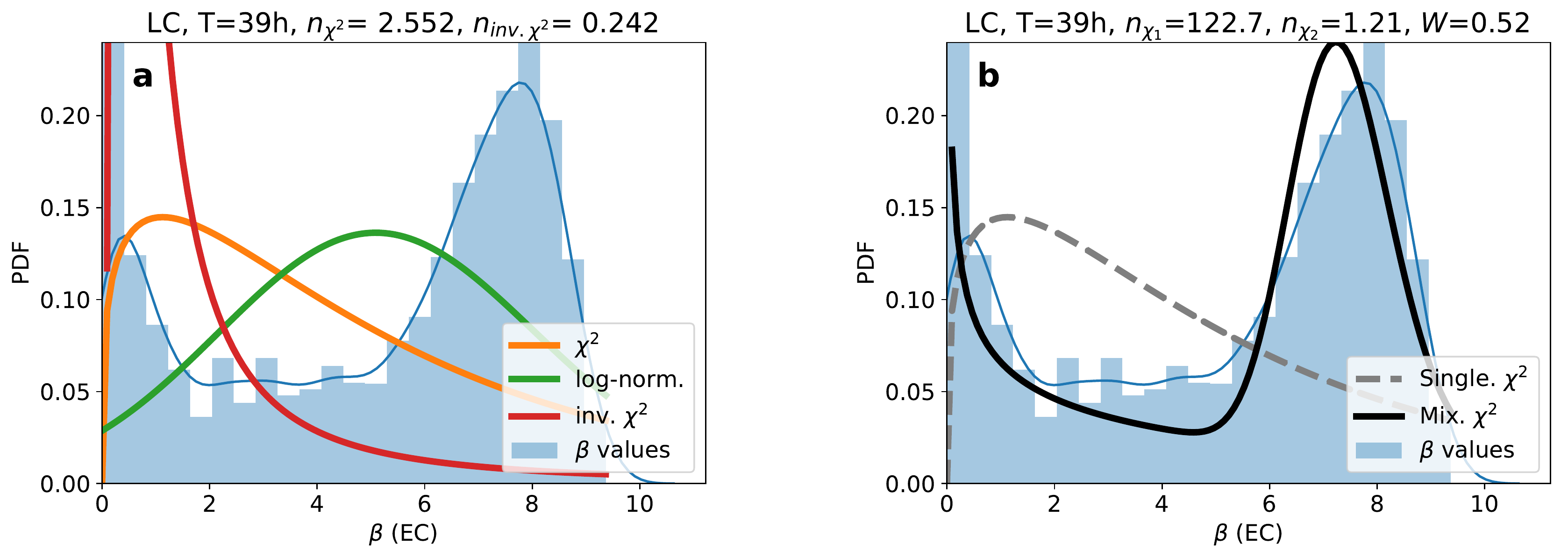}
\par\end{centering}
\caption{The extracted $\beta$-distribution of the electrical conductivity fluctuations does follow a mixture of two $\chi^2$-distributions. a: $\beta$-distribution with $\chi^2$, inv. $\chi^2$ and log-normal fit. b: $\beta$-distribution with  a single and the mixture $\chi^2$-distribution fitted.
Both plots use seasonal detrending at the LC measurement site.  The blue lines are Gaussian kernel estimates of the empirical PDF.
\label{fig:Beta_Mixture}}
\end{figure}

Indeed, this new mixture distribution of two $\chi^2$-distributions is an excellent fit to the data, see Fig. \ref{fig:Beta_Mixture} and Supplements for further examples.

\section*{Discussion}
In this paper we have shown that environmental time series
relevant for water quality in chalk rivers, such as the river Chess, behave in a superstatistical way.
We observe heavy-tailed distributions  for the aggregated statistics of oxygen and electrical conductivity. The dynamics of the measured time series is consistent with
that of a nonstationary process consisting of patches that locally exhibit
Gaussian behaviour, with the variance parameter fluctuating on a longer time scale $T$, which we extracted from the data. A new result is that the fluctuations of these water quality parameters do not follow Gaussian distributions as a whole but have distinct heavy tails that are well-approximated by $q$-Gaussian functions. This result is observed regardless of which detrending method (seasonal detrending and EMD) is applied. 
Using the average kurtosis, we determined the long time scale $T$ and found that the detrending method and specific detrending parameter only lead to linear scaling of the deduced long time scale, i.e. the superstatistical finding as such is robust with respect to the specific detrending method.
Consistent with the superstatistical assumptions, the local snapshots follow approximately Gaussian distributions and the $\beta$-distribution of oxygen fluctuations are approximated by log-normal distributions,  quite a similar
statistics as the one known for velocity and acceleration fluctuations in hydrodynamic turbulence.
 
An intriguing new finding is that electrical conductivity fluctuations at the LC site (contrary to oxygen fluctuations) display an unusual statistics,
namely a double-peaked $\beta$-distribution that is not immediately captured by existing superstatistical theory. We demonstrated how a $\chi^2$ mixture distribution can approximate the results but still, this finding points to the need of additional theoretical models that lead to double-peaked $\beta$-distributions. \revise{As a first step towards this extended theory, we propose a mixture $\chi^2$. Other possibilities to extend superstatistics could include  bivariate superstatistics \cite{caamano2020bivariate}.} 

Our superstatistical analysis requires the initial detrending of the data, illustrating that fluctuations of environmental time series are generally  not homogeneous in time. The data analysed here are somewhat comparable to other environmental time series with seasonal influence, e.g. the analysis of ambient temperature \cite{yalcin2013environmental}. Our approach could be applied to other seasonal time series: First, decompose the full time series into trend and fluctuations and then extract the distributions of the fluctuations as being heavy-tailed, followed by further superstatistical analysis to extract the relevant time scales and distributions of the parameter $\beta$. 

Interestingly, the impact of the sewage treatment works on the heavy-tail statistics is limited: Regardless of location, we did observe similar highly non-Gaussian distributions of the fluctuations, i.e. both upstream and downstream of the sewage treatment site (Fig. \ref{fig:Histograms_detrended}). Contrary, the long time scale, especially when using seasonal detrending on oxygen and EMD on electrical conductivity, displays qualitatively different behavior for the upstream and downstream locations (Fig. \ref{fig:Long_time_scale_scaling}), illustrating that human influence can be seen via time scale parameters extracted from the superstatistical analysis.
In particular, the long time scales diverges for lower filtering parameters at the two downstream locations with lower oxygen and higher electrical conductivity values. Meanwhile, we observe the  doubled-peaked $\beta$-distribution are most pronounced for the LC and BH sites (upstream). This might indicate that the double peaked $\beta$-distribution can emerge due to natural fluctuations, while fluctuations at the two downstream locations (LP and WB) might be further influenced by the human activity. Still, further research is necessary to fully understand this aspect.
\revise{Finally, we note that the observed  $q$-Gaussians imply a larger number of extreme events compared to any Gaussian process and the presented superstatistical approach provides a means to quantify this.}

A future project would be to compare our results obtained for the river Chess with the statistics generated by other
environmental time series, in particular comparing different rivers in a systematic and quantitative way or include other parameters \cite{kumar2011assessment}. Moreover, from a theoretical point of view, it would be desirable to expand the superstatistical theory relevant in nonequilibrium statistical physics towards double-peaked $\beta$-distributions, as these
distributions seem to appear naturally in the environmental context.

\begin{acknowledgements}
This project has received funding from the European Union’s Horizon 2020 research and innovation programme under the Marie Sklodowska-Curie grant agreement No 840825, from the Queen Mary University of London Centre for Public Engagement, and from Thames Water to install water quality sensors in the River Chess as part of the ChessWatch \cite{ChessWatch} project. This work would not have been possible without our Sensor Guardians who maintained the sensors in the river Chess throughout the monitoring period, and the landowners who gave us permission to install the sensors at each site.
\end{acknowledgements}

\subsection*{Author contributions}
B.S. \& C.B. conceived the research, B.S. generated all figures, K.H. \& H.R. collected and processed data, all authors contributed to writing the manuscript and interpreting the results.

\subsection*{Data and code availability}
\revise{Code to reproduce the results presented here is available at: 
\url{https://osf.io/mxcrv/}}

\subsection*{Declaration of interests}
The authors declare no competing interests.

\bibliographystyle{apsrev4-1}
\bibliography{references}

\clearpage

\appendix{Supplemental Material}

\begin{figure}[ht!]
\begin{centering}
\includegraphics[width=0.95\columnwidth]{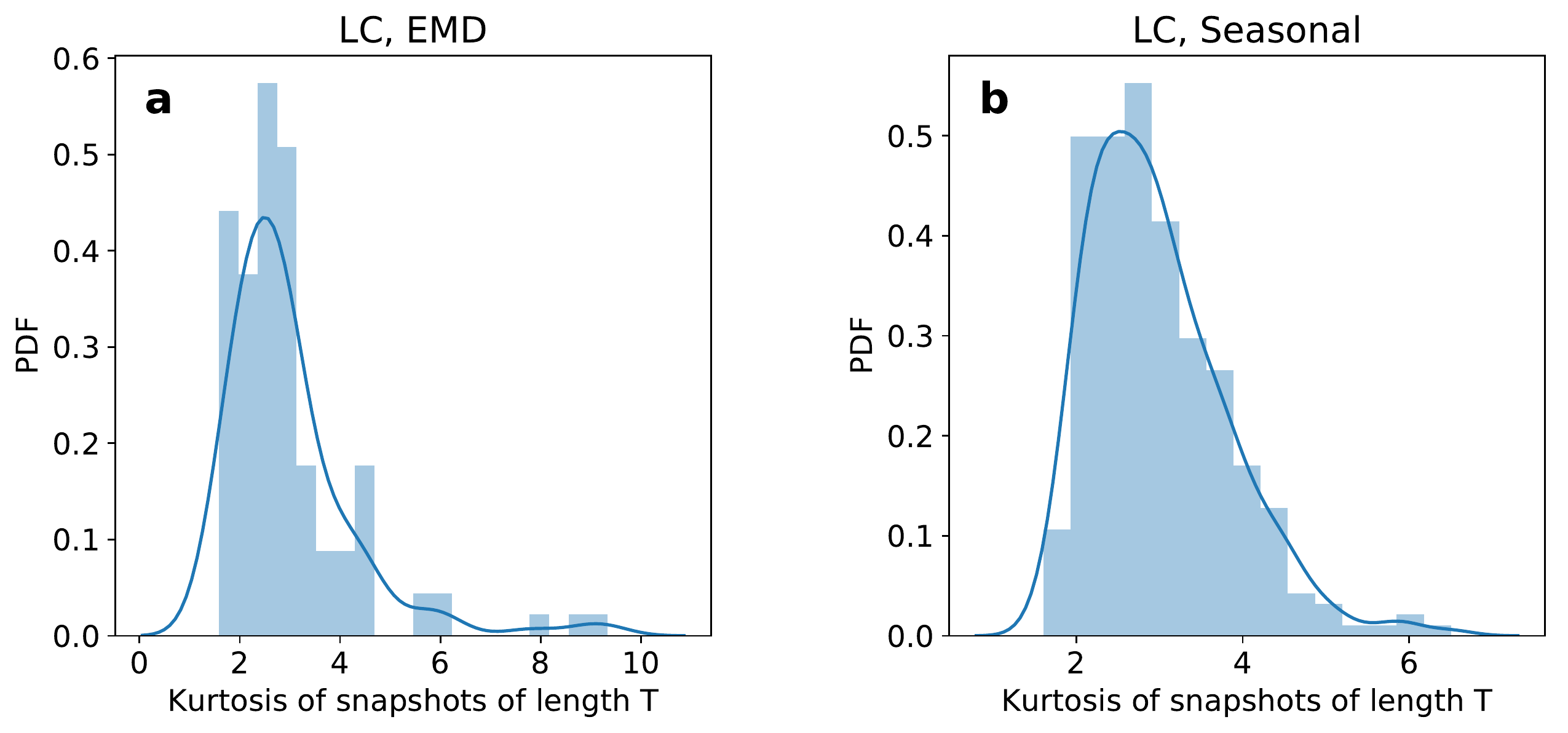}
\par\end{centering}
\caption{Kurtosis distribution of trajectory snapshots of length $T$. On average, we observe a Gaussian distribution but locally each distribution might be slightly leptokurtic or platykurtic, i.e. have a kurtosis larger or smaller than 3. \label{fig:Kurtosis_dis}}
\end{figure}

\begin{figure}[ht!]
\begin{centering}
\includegraphics[width=0.45\columnwidth]{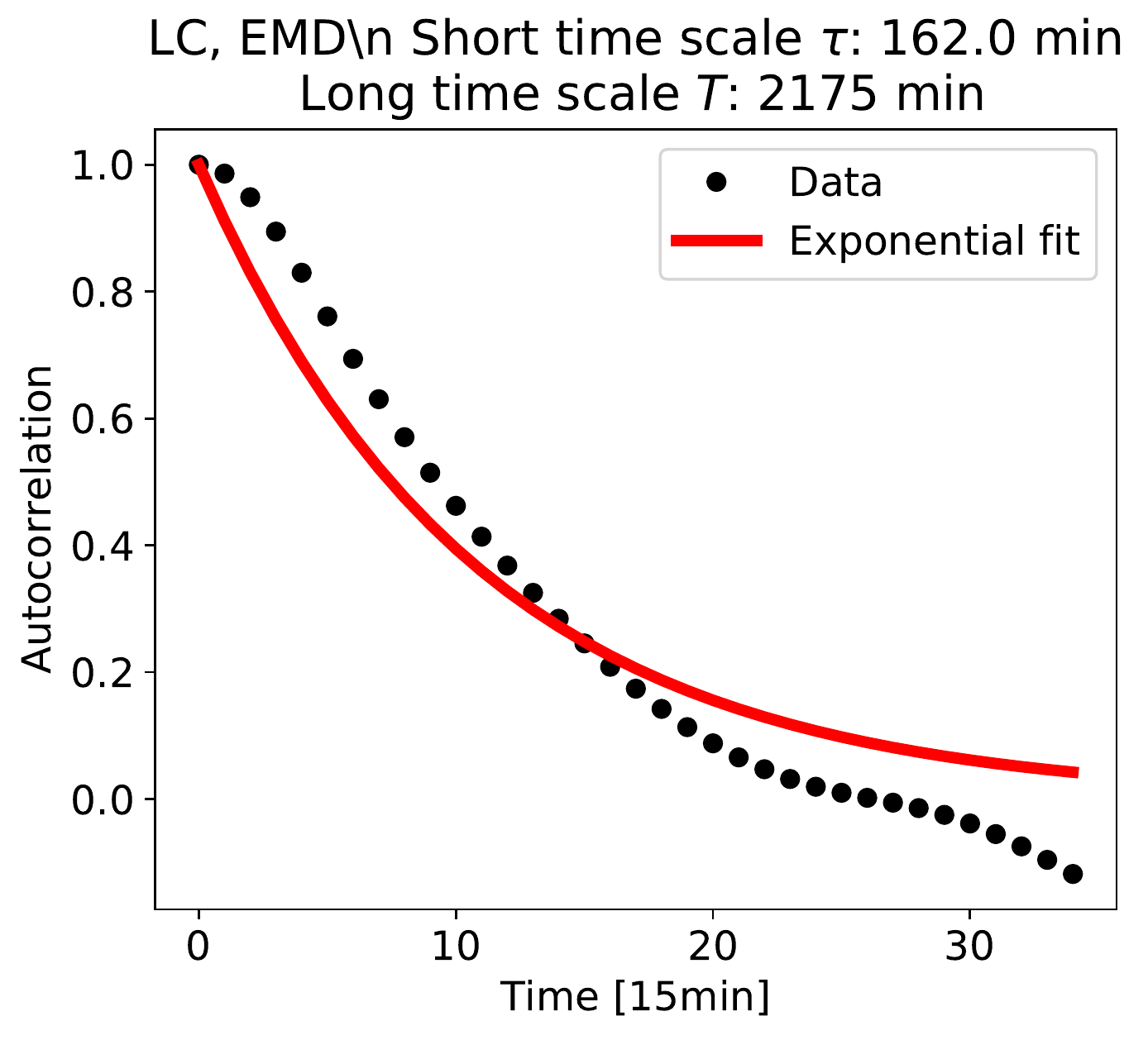}
\par\end{centering}
\caption{Time separation: The short time scale $\tau$, as determined by the approximately exponential autocorrelation decay $c(t) \sim \text{e}^{(-t/ \tau )}$ is at least one order of magnitude faster than the long time scale $T$. Here, we show this for the LC site, using EMD detrending when analysing dissolved oxygen. \label{fig:Shrt_time_scale}}
\end{figure}

\begin{figure}[ht!]
\begin{centering}
\includegraphics[width=0.6\columnwidth]{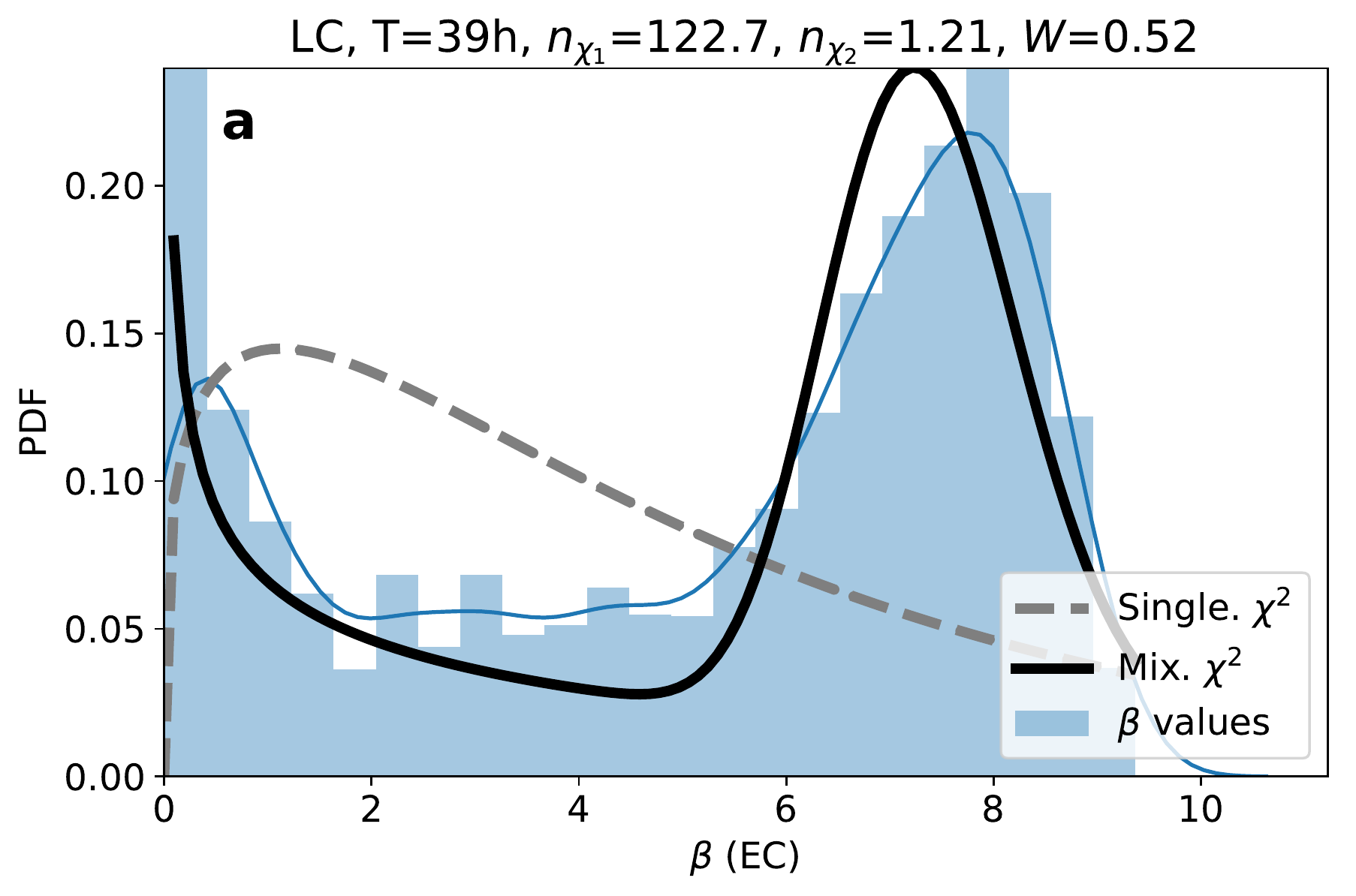}\\
\includegraphics[width=0.6\columnwidth]{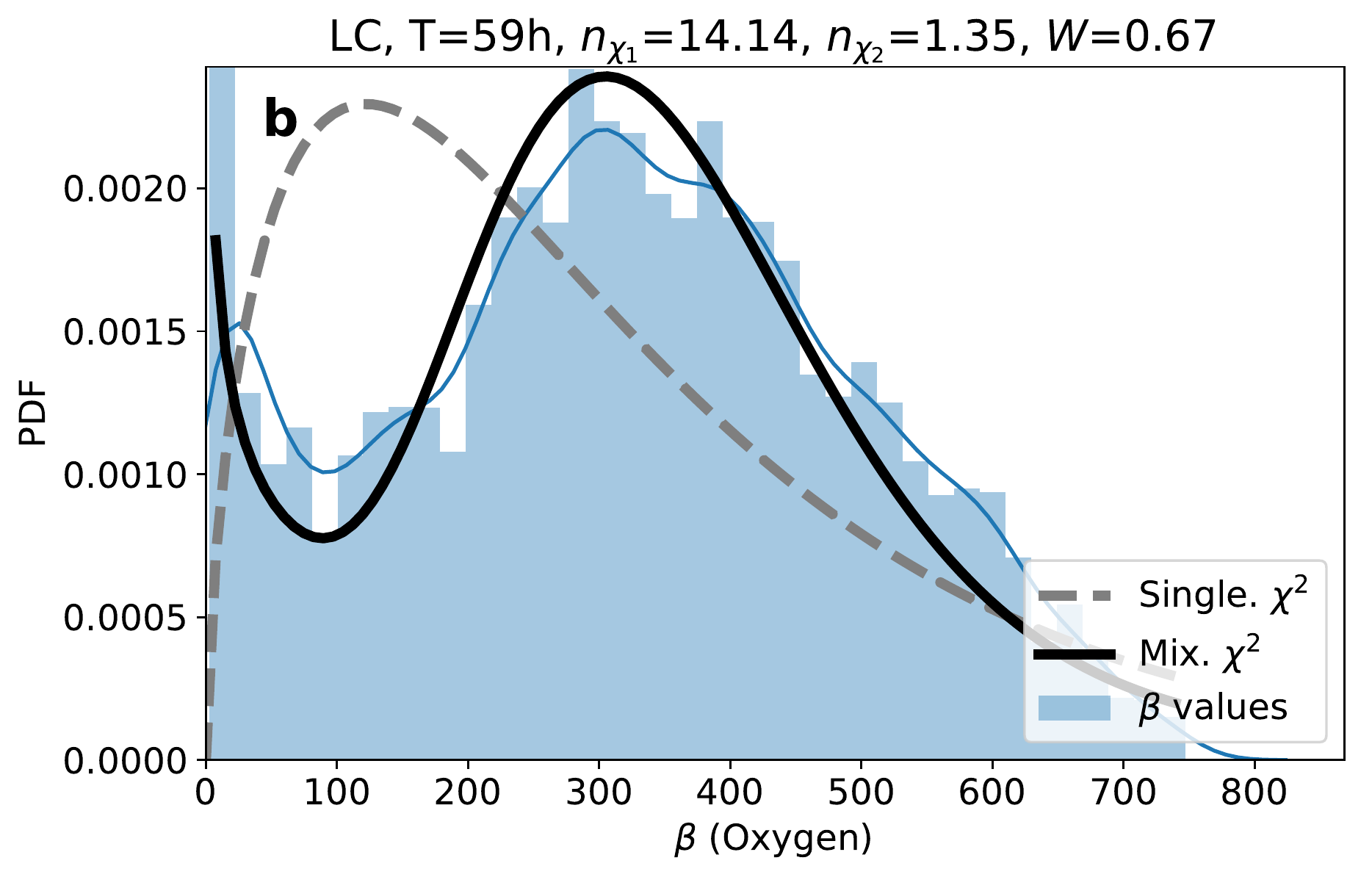}\\
\includegraphics[width=0.6\columnwidth]{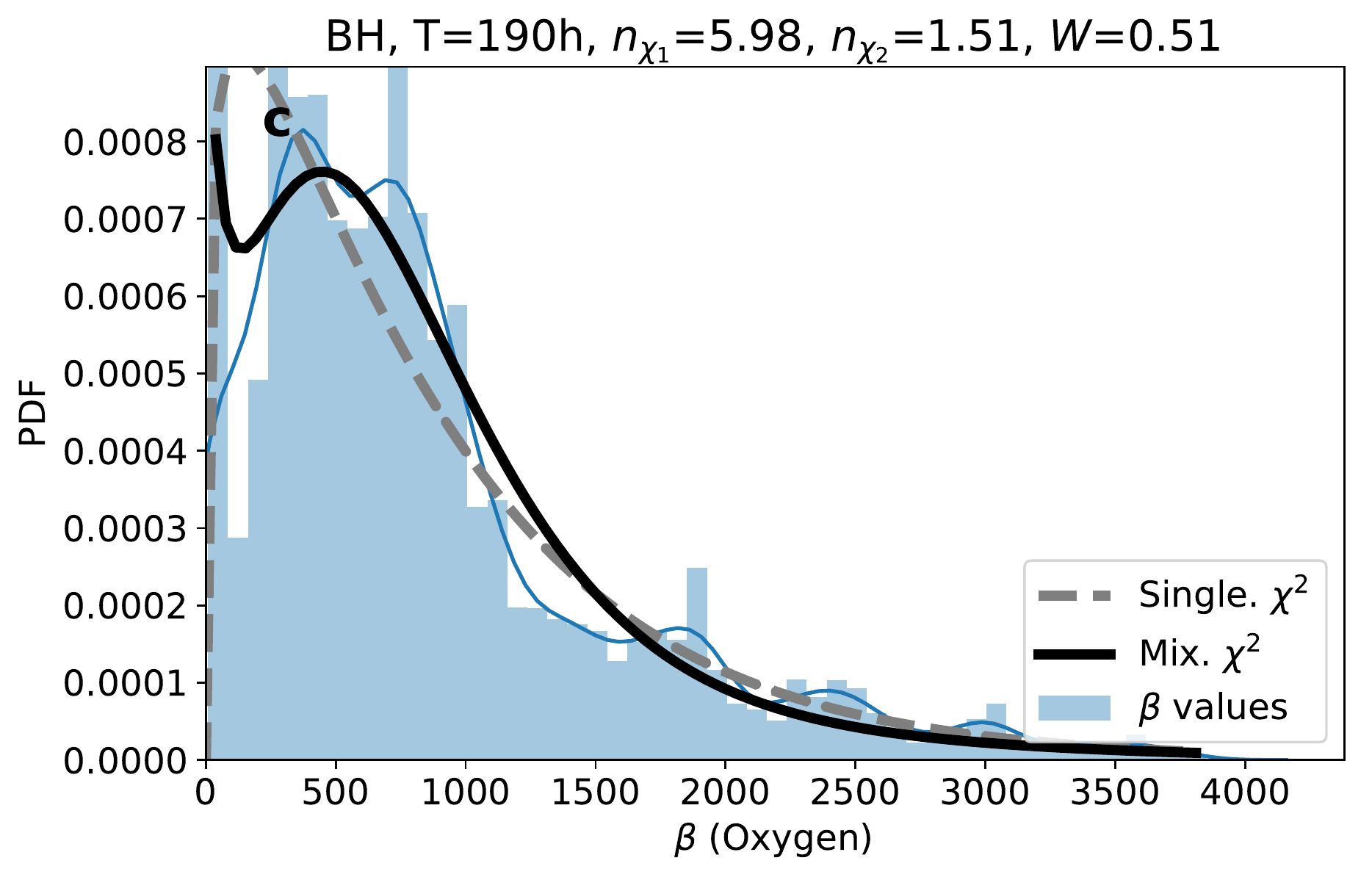}
\par\end{centering}
\caption{Examples of mixture $\chi^2$ distributions. a: Sesonal detrending of EC measurements at the LC site, b: Sesonal detrending of oxygen measurements at the LC site, c: EMD detrending of oxygen measurements at the BH site. The mixture $\beta$ distribution is given by $f(\beta)=W f_{\chi^2}(\beta, n_{\chi_1}, \beta_0) + (1-W) f_{\chi^2}(\beta, n_{\chi_2}, \beta_0)$. \label{fig:Mixtures}}
\end{figure}

\end{document}